\documentclass[aps,twocolumn]{revtex4}
\usepackage{graphics}
\usepackage{epsfig}
\usepackage{dcolumn}
\usepackage{amsmath,mathptm}
\usepackage[usenames]{color}

\begin{document}

\title{Discrete nonlinear Schr\"{o}dinger equation dynamics in complex networks}

\author{F. Perakis$^{1}$}
\author{G. P.  Tsironis$^{2}$}
\address{$^{1}$Physikalisch-Chemisches Institut, Universit\"at Z\"urich, Winterthurstrasse 190, CH-8057 Z\"urich, Switzerland}
\address{$^{2}$Department of Physics, University of Crete and Institute of Electronic Structure and
Laser, FORTH, P.O. Box 2208, Heraklion 71003, Crete, Greece.}

%\newpage

%\tableofcontents

%\newpage

\begin{abstract}

We investigate  dynamical aspects of the discrete nonlinear
Schr\"{o}dinger equation (DNLS) in finite lattices. Starting from
a periodic chain with nearest neighbor interactions,
 we insert randomly links connecting distant pairs of sites across
 the lattice. Using localized initial conditions we focus on the time averaged
  probability of occupation of the initial
 site as a function of the degree of complexity of the lattice and nonlinearity.
 We observe that selftrapping occurs at increasingly larger values of the nonlinearity
  parameter as the
 lattice connectivity increases, while close to the fully coupled network limit,
 localization becomes more preferred. For nonlinearity values
  above a certain threshold we
find a reentrant localization transition, viz. localization when
the number of long distant bonds is small followed by
delocalization and enhanced transport at intermediate bond numbers
 while close to the fully connected limit localization reappears.

\end{abstract}

\maketitle

%--------------------------------------------------------------------------------------------------------------------------------------------------------

\section{Introduction}
In this work we focus on  nonlinear aspects of dynamics in finite
discrete complex systems and investigate the simultaneous
presence of nonlinearity  with bond randomness and long range coupling.
Two basic features are employed, the first is selftrapping, viz.
a generic dynamical property of nonlinear lattices that leads to
localization in a translationally invariant system while   the
second is the geometrical feature that emerges from a small world
network (SWN).  In order to probe the former we use the  Discrete
nonlinear Schr\"{o}dinger  (DNLS) equation that is a ubiquitous
equation arising  in a variety of contexts  in condensed
matter~\cite{Hol59,Dav73,Scott85}, in optics~\cite{Led08}, in
Bose-Einstein condensation ~\cite{And98}, etc.  For the second  we
use Newman-Watts Small World Networks (SWN's) where random
connections are inserted linking distant pairs of lattice sites
originally linked through nearest neighbor coupling only
~\cite{Watts98,Watts99,Barabasi02}.  In a SWN there is coexistence
of disorder due to the random links, while the latter may at the
same time also enhance transport due to their long range
character.  The additional presence of nonlinearity introduces
further competition with the long range character of the network.
The coexistence of bond disorder, long range coupling and
nonlinearity makes the problem interesting to study.

The DNLS equation is given by:

\begin{eqnarray}
i\frac{d\psi_n}{dt}=\sum_{m} V_{n,m} \psi_{m}-\gamma |\psi_n|^{2} \psi_n
\label{dnls1}
\end{eqnarray}

where $\psi_n(t)$ is a complex amplitude with $|\psi_n|^{2}$
representing  the probability of finding an excitation at site
$n$ at time t, $V_{n,m}$ denotes the overlap integral between
sites $n$ and $m$ and $\gamma$ is a parameter that controls the
strength of the nonlinear term. In condensed matter context the
nonlinear term is due to polaronic effects in certain
semiclassical approximation~\cite{HT99}; we take the  total probability
normalized to unity $\sum|\psi_n|^2 = 1$. In applications in
optics, {$\psi_n(t)$} is the electric field in the $n$-th fiber
with $t$ denoting length, while  the overlap integral  and
nonlinear term are the evanescent field coupling term and Kerr
nonlinearity term respectively. Hereafter  we will
address the problem as semiclassical dynamics of a quantum
particle, keeping in mind that all results transfer to the
equivalent mathematical problems in optics, BEC, etc.

The matrix of coupling coefficients with elements $V_{n,m}$ is
taken to be a real symmetric matrix with diagonal elements
representing the local on-site energy; in the present work we will
take the latter equal to zero. In most studies one considers the
nearest-neighbor (NN) limit  with $V_{n,m}=V(
\delta_{n,m+1}+\delta_{n,m-1})$; in this case when $V <0$, $\chi
>0$ we have what is referred to as the focusing DNLS equation
while for $V <0$, $\chi < 0$ we have the defocusing one.  In the
general case we consider here, one may write $V_{n,m}=V \cdot
M_{n,m}$ with $M = [M_{nm} ]$ being the $N\times N$ adjacency
matrix with zero diagonal and off-diagonal elements either zero (when there is no
coupling) or one (when there is). Thus, the DNLS equation takes
equivalently the form~\cite{Scott85}

\begin{eqnarray}
i\frac{d\Psi}{d \tau}=M \Psi-\chi D(|\Psi|^{2}) \Psi
\label{dnls2}
\end{eqnarray}

where $\Psi = col(\psi_1,\psi_2,\dots,\psi_N)$ is an N-component
column vector, each component representing the complex
wave amplitude at a particular site, $D(|\Psi|^2) \equiv
diag(|\psi_1 |^2 ,|\psi_2 |^2,\dots,| \psi_N|^2)$ is a diagonal
$N\times N$ matrix representing the nonlinear term which
introduces anharmonicity. The parameter $\chi$ is the rescaled
nonlinearity parameter, viz. $\chi =\gamma /V$ while time has been
also rescaled to $\tau = V \cdot t$. We observe from Eq. (\ref{dnls1})
that reversing time is equivalent to solving the equation for
${\psi^*}_n(t)$, the complex conjugate of $\psi_n(t)$; the time
reversed equation results also when we change simultaneously the
signs of $\gamma$ and all $V_{n,m}$'s.  Thus, the single parameter
equation (\ref{dnls2}) for $\chi > 0$ is equivalent to the
defocusing case while for $\chi < 0$ we obtain the focusing case;
in both cases we should use proper initial conditions that may be
complex conjugates of the original ones.

\begin{figure} [h]
  \includegraphics{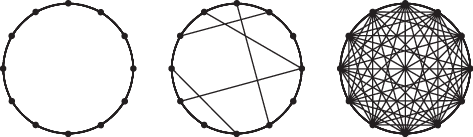}
  \caption{Starting from the nearest neighbor limit ($B=0$), we add randomly bonds until
the mean field limit ($B=B_{max}$).}
  \label{Fig1}
\end{figure}

One of the most interesting features of DNLS is that it leads to
self-trapping, in other words for small values of $\chi$ the
probability wave is delocalised while for values above a certain
treshold it localizes ~\cite{Scott85}. The self-trapping
properties of  DNLS for finite chains of sites have been studied
extensively in Ref ~\cite{Mol93}. But simple lattice models are
often insufficient to describe more complex systems encountered in
nature. In general such systems are characterized by networks with
bonds connecting sites with a wide distribution of mutual
distances. Examples can be found in various fields, ranging from
physics or biology to social science and computers. This
structural complexity gives rise to  interesting statistical
properties. These systems are known as small-world networks, i.e.
networks whose bonds possess a high degree of local
 clustering, and at the same time a short average path length connecting the nodes of
  the system ~\cite{Watts98,Watts99,Barabasi02}.

The study of the DNLS equation selftrapping properties in a simple one dimensional lattice
was presented in Ref.~\cite{Mol93}; in the present article we go beyond this work and
investigate selftrapping in
structurally complex networks. Specifically, we focus on the
dependence of localization properties of DNLS on the degree of
complexity of the network; the problem is set as follows: We begin
with N lattice sites that form a one dimensional ring with nearest
neighbor interactions only (Fig 1). For relatively large values of
N the DNLS dynamics on this ring is not different from the one in
very long one dimensional lattices~\cite{Mol93}.  Subsequently, we
destroy the periodicity of the lattice by randomly inserting
connections between distant sites ~\cite{Blum07,Quintanilla}. As
an initial condition we use that which places the probability
initially on one site of the system. The main quantity studied
throughout this paper is the long time-averaged probability $P
(\tau )$, viz. the occupation probability of the initially
populated site, taken to be the  site $n=1$:

\begin{eqnarray}
\left\langle P \right\rangle_{T} = \frac{1}{T} \sum_{\tau =0}^T |\psi_1 (\tau ) |^2
\label{e-av}
\end{eqnarray}
where $T$ is a long dimensionless time.
Physically, we place a quantum particle at a given lattice site
and follow its time evolution in the semiclassical limit. This
study is related to work done on DNLS in disordered
lattices~\cite{MT95,Shepel,Aubry}.  We note that other measures,
such as the inverse participation ratio, may be used in order to
probe the localization properties of the nonlinear network; we
verified numerically that the results to be presented below are
fully compatible with  both measures.

%---------------------------------------------------------------------

\section{Selftrapping in a complex  lattice}
Before we address the dynamics of the  DNLS equation in a small
world network we need first to establish two specific limits. In
order to form a small world network, we start from a periodic ring
of sites and  add randomly (using a uniform distribution)   $B$
bonds linking distant sites. For $B=0$  only first neighbors are
connected; this is the nearest neighbor (NN) limit. On the other
hand, the system is fully connected for the  maximum number of
possible distant connections, viz. for  $B_{max}=N \times
(N-1)/2-N $; we refer to this as the fully connected or mean field (MF) limit. In the
nearest neighbor limit we have the DNLS equation on a regular
lattice with periodic boundary conditions. In this case, when
nonlinearity is zero, an initially localized state spreads
throughout the lattice and after long times the probability to
be found at a given site will be of order $1/N$.
Increasing the nonlinearity parameter $\chi$ leads  to breather formation
signaled by selftrapping for values in the range $\chi \approx
 3-5$  ~\cite{Mol93}.  In the MF limit, on the other hand, the
 network is fully connected and this has ramifications for the DNLS
 selftrapping properties.  The purely linear chain on a small world
  network has been investigated in Ref~\cite{Blum07}, where it was found that partial localization of the probability at the initial
   site occurs for small $B$.

\begin{figure}[!ht]
  \includegraphics{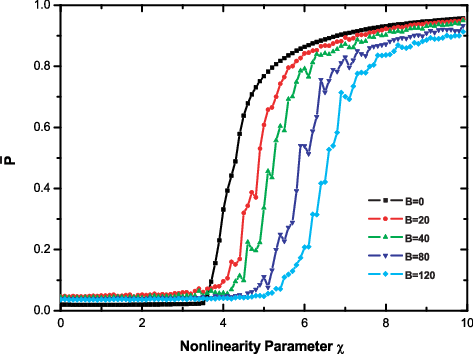}
  \caption{(Color online) Time and realization averaged probability $\bar{P}$  as a function of the nonlinearity parameter $\chi$ for different dilute SWN lattices.  Selftrapping occurs at increasingly larger values of $\chi$ as a function of $B$.}
  \label{Fig2}
\end{figure}

To study the general case of the DNLS equation on a SWN we place initially
all probability at a given site ($n=1$)  and for  each lattice we
follow the evolution of the probability amplitudes as a function
of time.  We evaluate the probability to return at the initial
site $|\psi_1 (T )|^2$, for long times ($T_{max} =10^5$) while
after time-averaging  we obtain $<P >_T$. This numerical
experiment is repeated for R distinct lattice realizations and,
finally, we obtain the averaged self-probability

\begin{eqnarray}
\bar{P}=\frac{1}{R} \sum_{R} <P>_{T}
\end{eqnarray}

 over these realizations. We present
results for $N=100$ where the ensemble averaging is done over
$R=500$ realizations. Repeating the above procedures for different
values of the nonlinearity parameter $\chi$ results in the
behavior seen in  Fig. (\ref{Fig2}).  We observe that the
known selftrapping transition of the finite, extended lattice
($B=0$) occurring for $\chi \simeq 3.5$~\cite{Mol93} shifts to
larger values of the nonlinearity parameter as we increase the
distant bond number $B$. In other words, the onset for the
breather regime close to the NN limit becomes delayed by the
presence of a small number of distant bonds. This delay is because,  the addition of few distant
bonds in this limit, even though adds disorder to the system, it also
increases the connectivity of the lattice and, as a result, the
amount of nonlinearity necessary for selftrapping is higher.  For negative nonlinearity parameters
the behavior in this regime is qualitatively similar.

%--------------------------------------------------------------------------------------------------------------------------------------------------------

In order to explore the whole range of complex lattices from $B=0$ to $B=B_{max}$ we first focus on the dynamics
of the two limiting lattices, viz. the NN and the  fully coupled MF limit; in the latter  every site is connect to all other sites.

\subsection{Nearest neighbor limit}
\begin{figure}[t]
  \includegraphics{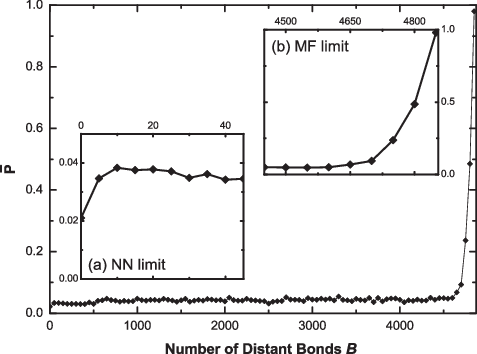}
  \caption{(Color online) Time and realization averaged probability of occupation of the initial site as a function of
the bond number $B$ for linear case. Insets: (b) detail of
the nearest neighbor regime and (c) detail of the mean field regime.}
  \label{Fig3}
\end{figure}

In the NN limit an  initially localized state will propagate
throughout the lattice unless the nonlinearity parameter is strong
enough; in this case selftrapping will occur and only part of the
probability will escape the initial site~\cite{Mol93}.   In the
linear problem  the authors or ref. ~\cite{Blum07} showed that the addition of
a small number of SWN bonds leads to faster delocalization with
the appearance of very weak localization.  We probe this regime also for the nonlinear
case and present the results in Fig. \ref{Fig3} for the linear case, and in Fig.\ref{Fig4} and \ref{Fig5} for the nonlinear case. We find that in this
regime small nonlinearity does not modify the conclusions of ref.
\cite{Blum07}, viz. delocalization persists while some remnant weak
localization survives.  When, however nonlinearity surpasses a
certain value, nonlinear localization appears that dominates the
dynamics.

\subsection{ Mean field limit}
In the purely linear case ($\chi =0$) the nonlinear eigenvalue problem of Eq. (\ref{dnls2}) reduces to
\begin{eqnarray}
M\Psi=\lambda \Psi \Rightarrow det[M-\lambda I] = 0
\label{e-lin}
\end{eqnarray}
where $\Psi$ is the N-component column vector and $M$ is the
adjacency matrix that contains both NN and long range couplings.
This problem can be diagonalized straightforwardly observing that
the matrix $M$ can be decomposed to $M=K-I$ where $K$ is an $NxN$
matrix with all elements identical to unity while $I$ the
corresponding unit matrix. The matrix $K$  has rank one and
obvious eigenvector $(1 1 1 ... 1)^T$ with eigenvalue $N$. Since
$Tr(K) =N$, all other $N-1$ eigenvalues are identically equal to
zero.  The matrix that diagonalizes $K$, diagonalizes $M$ as well,
resulting in a spectrum for $M$ consisting of a $N-1$ degenerate
eigenstates with value $-1$ and one (Peron-Frobenius) eigenvalue
with value $N-1$. If the eigenvectors of $M$  have the form $(a_1~
a_2 ~ a_3 ~... a_N )^T$, then  the one corresponding to the
 eigenvalue $\lambda_{PF}=N-1$ has $a_1 = a_2 =
...=a_N = 1$ with normalization $1/\sqrt{N}$ while the other $N-1$
eigenvectors ( that project to the kernel of K) are solutions of the
equation $\sum_{i=1}^N a_i =0$ with $\sum_{i=1}^N {a_i}^2 =1$. We
observe that as the total number of sites $N$ increases the
degeneracy of the system increases while the delocalized
 state eigenvalue $\lambda_{PF}=N-1$ becomes singular.  Indeed,  in the infinite
lattice limit the band in $k$-space becomes optical with one
singular value for the long wavelength mode at $k=0$.

After some straightforward matrix manipulations and assuming that
the particle is initially placed at site $n=1$, viz. $\psi_1 (0)
=1$ we obtain the following time evolution for the lattice
occupation probability  in the linear MF limit:
\begin{eqnarray}
|\psi_1 (\tau )|^2 =\frac{1}{N^2}\left[ 1+(N-1)^2 + 2(N-1)\cos(N \tau ) \right]  \\
|\psi_m (\tau )|^2 =\left[ \frac{2 \sin (N\tau /2 )}{N} \right]^2 , ~~~m\ne 1
\end{eqnarray}
In the limit $N \rightarrow \infty$  we have $ |\psi_1 (\tau )|^2
=1$ and $|\psi_m (\tau )|^2 =0$, for $m\ne1$.  We observe that for
large $N$ in the MF limit we have extreme localization stemming
from the  degeneracy of the lattice eigenvalues.  Only when the
particle is spread initially equally on all sites it may stay in
this delocalized state  corresponding to the Perron-Frobenius
eigenstate of the lattice. We thus find that in the linear case
while the NN limit favors delocalization the MF limit promotes
localization~\cite{Bor99,Naz99}.  This feature will  dominate the
behavior of DNLS in the large-$B$ limit.

The nonlinear MF problem for localized initial condition can be
also solved analytically ~\cite{GPT}; we find that for positive
nonlinearity parameter the increase of $\chi$ leads to further
localization, enhancing thus the localizing tendency introduced by
the linear MF case.  On the other hand, for negative nonlinearity
parameter we observe detraping followed subsequently by retrapping
accompanied by change of character of the solution. The resulting
transitions, however, occur for values of nonlinearity that are of
order $N$ and thus do not play any direct role in the present
study; details on the nature of the dynamics in this limit will be
presented elsewhere.

\section{From the nearest neighbor  to the fully connected limit}
\begin{figure}[t]
  \includegraphics{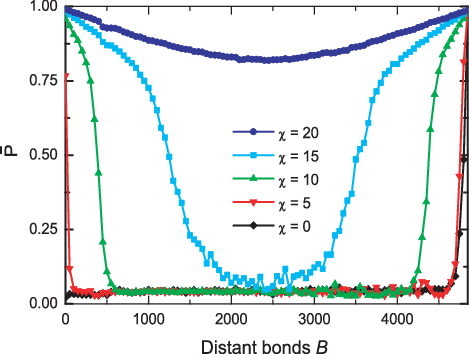}
  \caption{(Color online) Time and realization averaged probability of occupation of the initial site as a function of
the bond number $B$ for different positive nonlinearity values.}
  \label{Fig4}
\end{figure}

\begin{figure}[t]
  \includegraphics{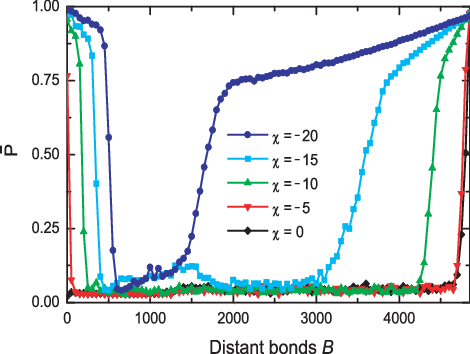}
  \caption{(Color online) Time and realization averaged probability of occupation of the initial site as a function of the bond number $B$ for different negative nonlinearity values.}
  \label{Fig5}
\end{figure}

In order to study the complete SWN problem
we construct the lattice by starting from the NN limit where $B=0$
and  randomly adding bonds  until we reach the MF  limit where
$B=B_{max}$ (Fig. \ref{Fig1}). We use $N=100$ and thus
$B_{max}=4850$.  The  results for the the whole range from the NN to
the MF limits are presented  in Figs. (\ref{Fig3}), (\ref{Fig4}) and (\ref{Fig5}) .

For positive nonlinearities, while $\chi$ is very small, the delocalization tendency
introduced by the long
 distant bonds prevails.  This is seen in Fig. (\ref{Fig2}) where for nonlinearity values smaller than
 approximately $3.5$ non localization appears.  However, for larger nonlinearities, the initially placed
 probability at site $1$ becomes selftrapped, leading $\bar{P}$ values close to unity in the NN limit; this is
 seen in Fig. (\ref{Fig4}).  As the bond number increases from $B=0$ to small values we observe a precipitous drop in
 $\bar{P}$ in the small $B$-limit signifying that the addition of distant bonds overcomes the localization induced
 by nonlinearity.  We note, however, that as nonlinearity increases
 the  delocalization induced by the distant bonds becomes less potent.  As the number of bonds increases we observe a reversal
 of the tendency, i.e.  the increase of $B$ induces localization.  While nonlinearity is small,  this
 relocalization occurs relatively abruptly and only in the zone very near to $B_{max}$.  For larger nonlinearities,
 e.g for  $\chi =20$, this transition to localization occurs for $B\gtrsim B_{max}/2$ and
 is less sharp.

 As nonlinearity increases, the transition to localization in the MF limit occurs for smaller bond number $B$; this
 signifies that this transition is primarily driven by the localization tendency of the linear MF model, augmented however
 by the presence of nonlinearity.  We thus find that both close to the NN and MF limits the nonlinear model shows additional
 localization tendency induced by the  nonlinearity parameter while for intermediate values of $B$ we typically observe
 delocalization.  The numerical results for negative $\chi$ are presented in Fig. (\ref{Fig4}) and show similar behavior  except that the
 localization-delocalization transition occurs for smaller number of SWN bonds and is quite sharper than the
 corresponding one for $\chi > 0$.  As the value of $\chi$ increases in absolute value, additional differences from the positive-$\chi$ case appear including the more asymmetrical aspect of the localization-delocalization-relocalization transitions.

%--------------------------------------------------------------------------------------------------------------------------------------------------------

\section{Conclusions}

We investigated dynamical properties of the DNLS equation related
to selftrapping and the onset of nonlinear localization in complex
lattices in the form of small world networks. The nonlinearity
parameter $\chi$ and the number of small world connections $B$ are
the two variables that induce localization and delocalization
respectively.  We used as a basic quantity the probability to
return to the origin and observed that in the fully linear regime
($\chi =0$) the increase of the small world link number leads
essentially to delocalization due to the effective increase of the
lattice dimensionality.  This trend is generally true with the
exception for values very close to the mean field limit; in the
latter, the exact solution shows that the lattice exhibits complete
localization tendency for an initially localized state while, on the other hand,
the complete delocalized state is
an eigenstate of the lattice. We note, however, that in the very small $B$
regime, while the system is delocalized, some partial localization
is observed~\cite{Blum07}. When the nonlinearity parameter is non
zero we observe a competition between localization induced by the
latter and delocalization induced by the distant connections. This
competition makes selftrapping occur at larger $\chi$-values
compared to the translationally invariant lattice case while for
very large nonlinearities the probability becomes to a large
extent selftrapped.

An interesting feature appears from the linear MF behavior, viz.
close to the large $B$ regime the nonlinear localization tendency
is generally augmented by the linear MF localization behavior.
Thus for nonlinearity parameter values above a certain threshold
we observe as a function of long range bond number first
localization (in the small $B$ range), delocalization at
intermediate numbers of links and relocalization close to to the
MF limit.   The occurrence of this reentrant transition is due to
two different localizing factors, nonlinearity close to the NN
limit and  state degeneracy in the linear MF limit. The DNLS
properties on the SWN may have some applications in charge and
energy propagation and storage in complex systems such as proteins
as well as optical lattices. One may envision flexible polymeric
channels that due to their variable distance may form a small
world network and thus lead to the properties presented here.

\end{document}